\documentclass[pra,twocolumn,amsmath,amssymb,nofootinbib]{revtex4}

\usepackage{graphicx}

\begin{document}
\title{Necessary and sufficient criterion for extremal quantum correlations\\
in the simplest Bell scenario}
\author{Satoshi Ishizaka}
\affiliation{Graduate School of Integrated Arts and Sciences,
Hiroshima University,
1-7-1 Kagamiyama, Higashi-Hiroshima 739-8521, Japan}
\date{\today}
%
\begin{abstract}
In the study of quantum nonlocality, one obstacle is that the analytical
criterion for identifying the boundaries between quantum and postquantum
correlations has not yet been given, even in the simplest Bell scenario. We
propose a plausible, analytical, necessary and sufficient condition ensuring
that a nonlocal quantum correlation in the simplest scenario is an extremal
boundary point. Our extremality condition amounts to certifying an
information-theoretical quantity; the probability of guessing a measurement
outcome of a distant party optimized using any quantum instrument. We show
that this quantity can be upper and lower bounded from any correlation in a
device-independent way, and we use numerical calculations to confirm that
coincidence of the upper and lower bounds appears to be necessary and
sufficient for the extremality.
\end{abstract}

%
\maketitle
%
Since Einstein, Podolsky and Rosen proposed a paradox \cite{Einstein35a} in
1935, quantum nonlocality has been a central topic in fundamental science. In
1964, Bell showed that the nonlocal correlations predicted by
quantum mechanics are inconsistent with local realism \cite{Bell64a}. The
nonlocal correlations do not contradict the no-signaling principle, but it was
later found that the strength of quantum correlations is more restricted than
that allowed by the no-signaling principle \cite{Tsirelson80a,Popescu94a}.
Since then, many efforts have been made to determine the fundamental principles
limiting quantum nonlocality \cite{Brunner14a,Popescu14a,Chiribella16a,Oas16a}.
In these studies, however, one serious obstacle is that the analytical
criterion for identifying the boundaries between quantum and postquantum
correlations has not yet been given, even in the simplest Bell scenario.

In the simplest Bell scenario, where two remote parties,  Alice and Bob, each
perform two binary measurements on a shared quantum state, Tsirelson showed
that the Bell inequality of the Clauser-Horne-Shimony-Holt (CHSH)
type \cite{Clauser69a} is violated up to $2\sqrt{2}$ by quantum
correlations \cite{Tsirelson80a}. The correlation attaining the Tsirelson bound
is an extremal point of the convex set of quantum correlations. When 
marginal probabilities of obtaining the
measurement outcomes are unbiased, the boundaries are identified
using the Tsirelson-Landau-Masanes (TLM) analytical
criterion \cite{Tsirelson87a,Landau88a,Masanes03a}.
In a general case where marginals may be biased,
several methods work for identifying
the boundaries, such as the Navascu\'es-Pironio-Ac\'{\i}n (NPA)
hierarchy \cite{Navascues07a,Navascues08a}, the see-saw iteration
algorithm \cite{Werner01b,Vertesi14a}, and the quantifier elimination
algorithm \cite{Wolfe12a}, but obtaining the analytical criterion is a
long-standing open problem. Recently, it was shown that the geometry of
the quantum set has rich and counterintuitive features \cite{Goh18a};
specifically, contrary to the case of unbiased marginals, flat
(i.e., non-extremal) boundaries made from nonlocal correlations exist (other
than the edges of the probability space), which
indeed tells us the difficulty of the problem. Therefore, it is reasonable and
worthwhile to determine the analytical criterion for identifying extremal
points, instead of full boundaries, as the quantum set is a convex hull of
extremal points.

In this paper, we propose a plausible, analytical, necessary and sufficient
condition ensuring that a nonlocal quantum correlation is extremal.
To this end, we focus on the optimal probability of guessing a measurement
outcome of a distant party, which was shown to play a crucial role in
constraining quantum correlations \cite{Ishizaka17a}.
We show that the guessing probability can be upper and lower bounded from any
correlation in a device-independent way, and as a result when the upper and
lower bounds coincide, the guessing probability can be certified (i.e.,
uniquely determined irrespective of details when realizing a correlation). 
We use numerical calculations to confirm that this certifiability condition
appears to be necessary and sufficient for the extremality.

To begin with, let us briefly summarize preliminaries. For details,
see \cite{Brunner14a} and the references therein. In the simplest Bell
scenario, Alice (Bob) performs a measurement on a shared state depending on
a given bit $x$ ($y$) and obtains an outcome $a\!=\!\pm 1$ ($b\!=\!\pm1$). We
can assume without loss of generality that they perform projective measurements
on a pure state $|\psi\rangle$. The properties of a nonlocal correlation are
described by a set of conditional probabilities $\mathbf{p}\!=\!\{p(ab|xy)\}$.
The set $\mathbf{p}$ specifies a point in the probability space, whereas a Bell
inequality, which has the form $\sum_{abxy}V_{abxy}p(ab|xy)\!\le\!c$, specifies
a hyperplane in the probability space. The left-hand side of the Bell
inequality is called the Bell expression. Let $A_x$ ($B_y$)
be the observable of Alice's (Bob's) projective measurements, which satisfy
$A_{x}^{2}\!=\!B_{x}^{2}\!=\!I$. Due to the no-signaling
condition such that the marginal $p(a|x)$ and $p(b|y)$ does not depend
on $y$ and $x$, respectively, all Bell expression can be recast, without loss
of generality, in the form
\begin{equation}
B=\sum_{x} V^{A}_{x} \langle A_x\rangle
+\sum_{y} V^{B}_{y} \langle B_y\rangle
+\sum_{xy} V_{xy} \langle A_x B_y\rangle,
\label{eq: Bell expression}
\end{equation}
where $\langle \cdots \rangle$ is the abbreviation of
$\langle\psi| \cdots |\psi\rangle$. The set of local correlations is
tightly enclosed by facet Bell inequalities, all of which are the CHSH type,
together with the positivity constraints $p(ab|xy)\!\ge\!0$. As a result,
\begin{equation}
C_{\rm CHSH}=\max |\langle A_0B_0\rangle+\langle A_0B_1\rangle+
\langle A_1B_0\rangle-\langle A_1B_1\rangle|
\label{eq: CCHSH}
\end{equation}
exceeds 2 if and only if the correlation is nonlocal, where the
maximization is taken over the four positions of the minus sign in the 
CHSH expression. A local correlation is an extremal point of the quantum set
if and only if it is a deterministic correlation. In this paper, therefore,
we exclusively consider extremal correlations made from a
nonlocal quantum correlation.

Let us then recall the bound on nonlocality in terms of the guessing
probability \cite{Ishizaka17a}, i.e.,
\begin{equation}
\sum_{xy}s_x u_{xy}(-1)^{xy}\langle A_x B_y\rangle
\le\big[\sum_{x}s^{2}_x (D^{B}_x)^2 \big]^{1/2}.
\label{eq: CQB}
\end{equation}
Any quantum realization must satisfy this inequality for any
real $s_x$ and $u_{xy}$ such that $u_{00}u_{01}\!=\!u_{10}u_{11}$ and
$\sum_{xy}u^{2}_{xy}\!=\!1$. The quantity $D^{B}_x$ describes the guessing
probability; Bob's optimal probability of guessing Alice's outcome $a$
is $(1\!+\!D^{B}_x)/2$ (when Bob's conditional states are pure).
The precise definition is given by Eq.\ (\ref{eq: Definition of D}) below.
The necessary and sufficient condition for the fulfillment of
Eq.\ (\ref{eq: CQB}) and of the complement inequality in terms of Alice's
guessing probability ($\forall s_x,u_{xy}$) is 
\begin{eqnarray}
\left|\tilde C_{00}\tilde C_{01}-\tilde C_{10}\tilde C_{11}\right|
&\le& (1-\tilde C^{2}_{00})^{1/2}
(1-\tilde C^{2}_{01})^{1/2} \cr
&&+(1-\tilde C^{2}_{10})^{1/2}(1-\tilde C^{2}_{11})^{1/2}
\label{eq: Condition 1}
\end{eqnarray}
for both $\tilde C_{xy}\!=\!\langle A_xB_y\rangle/D^{B}_x$ and
$\tilde C_{xy}\!=\!\langle A_xB_y\rangle/D^{A}_y$ \cite{Ishizaka17a}.
When $\tilde C_{xy}\!=\!\langle A_xB_y\rangle$, 
Eq.\ (\ref{eq: Condition 1}) reproduces the TLM inequality, and
the saturation is necessary and sufficient for the extremality of
{\it nonlocal} quantum correlations in the case of unbiased marginals
($\langle A_x\rangle\!=\!\langle B_y\rangle\!=\!0$) \cite{Wang16a,Goh18a}.
Therefore, Eq.\ (\ref{eq: Condition 1}) is said to be the scaled TLM
inequality, {\it as the correlation function $\langle A_x B_y\rangle$ is
scaled by $D^{B}_x$ and $D^{A}_y$}. As preliminarily mentioned
in \cite{Ishizaka17a}, every extremal correlation including the case
of biased marginals appears to saturate the scaled TLM inequality, whose
numerical evidence is explicitly shown later. However, it was also shown that
the saturation alone is insufficient for identifying the extremality.

To search for a complete set of conditions, let us focus on the fact that, for
a given $\{\langle A_xB_y\rangle,\langle A_x\rangle,\langle B_y\rangle\}$,
the upper bounds of $D^{B}_x$ and $D^{A}_y$ can also be determined irrespective
of the details of the realizations. This can be done by using the method based
on the NPA hierarchy \cite{Navascues07a,Navascues08a} as follows: Let us
consider the states of
$|A_x\rangle\!\equiv\!A_x|\psi\rangle$ and
$|B_y\rangle\!\equiv\!B_y|\psi\rangle$.
In addition, $|X\rangle\!\equiv\!X|\psi\rangle$ is introduced to obtain the
bound of $D^{B}_x$,
where $X$ is any Hermitian operator on Bob's side that satisfies
$\langle X|X\rangle\!=\!1$.
The Gram matrix $\Gamma$ of the
states $\{|\psi\rangle,|A_x\rangle,|B_y\rangle,|X\rangle\}$ has the form
\begin{equation}
\Gamma= \left(\begin{array}{cccccc}
 1 & \langle A_0\rangle & \langle A_1\rangle & \langle B_0\rangle & \langle B_1\rangle & \gamma_{16}  \\
 & 1 & \gamma_{23} & \langle A_0B_0\rangle & \langle A_0B_1\rangle  & \gamma_{26}  \\
 & & 1 & \langle A_1B_0\rangle & \langle A_1B_1\rangle & \gamma_{36} \\
 & & & 1 & \gamma_{45} & \gamma_{46} \\
 & & & & 1 & \gamma_{56} \\
 & & & & & 1 \\
\end{array}\right)
\end{equation}
where only the upper triangular part is shown. Since
\begin{equation}
D^{B}_{x}=\max_{\langle\psi| X^2|\psi\rangle=1}\langle\psi|A_x X|\psi\rangle
=\max_{\langle X|X\rangle=1} \langle A_x|X\rangle,
\label{eq: Definition of D}
\end{equation}
the upper bound of $D^{B}_0$ ($D^{B}_1$) is obtained by
maximizing $\gamma_{26}$ ($\gamma_{36}$) under the constraint
that the real symmetric matrix $\Gamma$ is positive semidefinite.
Here, the maximizations of $\gamma_{26}$ and 
$\gamma_{36}$ are done separately.
However, this method, corresponding to the lowest level of the NPA hierarchy,
does not work well, as $|X\rangle\!=\!|A_x\rangle$ is always a solution
and the maximum cannot be less than 1. Let us then
move on to the next $1\!+\!AB$ level of the NPA hierarchy.
Namely, let us further introduce
$A_x B_y|\psi\rangle$ and $A_x X|\psi\rangle$, and construct
the $12\!\times\!12$ Gram matrix (with constraints between
the matrix elements). Then, the bounds less than 1 can be obtained
(mainly numerically, though). The upper bounds of $D^{A}_y$ are obtained in the
same way. Throughout this paper, these bounds are called device-independent
upper bounds.

Now, consider a realization such that $D^{B}_x$ and $D^{A}_y$
coincide with the device-independent upper bounds,
and further saturates Eq.\ (\ref{eq: CQB}) for an
appropriate choice of $s_x$ and $u_{xy}$, hence saturating
Eq.\ (\ref{eq: Condition 1}). Such a correlation has a
significant property: $D^{B}_x$ and $D^{A}_y$ are unique irrespective of the
realizations, as they are tightly bounded from above and below. Namely,
they can be certified from
$\{\langle A_xB_y\rangle,\langle A_x\rangle,\langle B_y\rangle\}$,
as of the certification of, e.g., randomness \cite{Pironio10a}.
Note that, even in this time, $B_y$ itself generally does not coincide with
an optimal operator for guessing the outcome of $A_x$, and
vice versa (see Appendix C in \cite{Ishizaka17a}).
This certifiability of $D^{B}_x$ and $D^{A}_y$, despite that they
depend on the state and the measurements, may implicitly imply that the
realization is unique up to local isometry, i.e., the realization can be
self-tested \cite{Mayers04a}, as in the case of unbiased marginals where
every nonlocal boundary correlation self-tests the maximally entangled
state \cite{Wang16a}. Therefore, such a correlation is a good candidate of
an extremal correlation, as a correlation must be extremal if it is
self-testable \cite{Goh18a}. Moreover, if this insight is true,
the certifiability of $D^{B}_x$ and $D^{A}_y$ ensures that the
device-independent bounds are attained by a two-qubit realization,
as every extremal correlation in the simplest Bell scenario has a two-qubit
realization \cite{Tsirelson80a,Masanes06a}.

In two-qubit realizations, where projective measurements of rank 1 are
performed on a two-qubit entangled pure state
$|\psi\rangle\!=\!\cos\chi|00\rangle\!+\!\sin\chi|11\rangle$,
since the guessing probability is given by
$D^{B}_x\!=\!\hbox{tr}|\rho^{B}_{1|x}\!-\!\rho^{B}_{-1|x}|$ \cite{Helstrom69a},
with $\rho^{B}_{a|x}$ being Bob's local state conditioned on $a$,
and similarly for $D^{A}_y$, we have
(see Appendix \ref{sec: Two-qubit realization} for details)
\begin{equation}
(D^{B}_x)^2=\langle A_x\rangle^2+\sin^2 2\chi,\hbox{~}
(D^{A}_y)^2=\langle B_y\rangle^2+\sin^2 2\chi.
\label{eq: Dx and Dy}
\end{equation}
It is then found that,
for a given $\{\langle A_xB_y\rangle,\langle A_x\rangle,\langle B_y\rangle\}$,
the entanglement of $|\psi\rangle$ specified by $\sin^2 2\chi$
is determined as a consistent solution of four quadratic equations to be
\begin{eqnarray}
S^{\pm}_{xy}&\equiv&\frac{1}{2}\left[J_{xy}
\pm\sqrt{J^{2}_{xy}-4K^{2}_{xy}}\right], \cr
J_{xy}&\equiv&\langle A_xB_y\rangle^2-\langle A_x\rangle^2-\langle B_y\rangle^2+1, \cr
K_{xy}&\equiv&\langle A_xB_y\rangle-\langle A_x\rangle\langle B_y\rangle.
\end{eqnarray}
For each $x$ and $y$, one of the two solutions
$S^{\pm}_{xy}$ agrees with $\sin^2 2\chi$.
Since $D^{B}_x$ and $D^{A}_y$ are increasing functions of $\sin^2 2\chi$
as in Eq.\ (\ref{eq: Dx and Dy}), we immediately obtain the following
analytical upper bounds in two-qubit realizations:
\begin{equation}
(D^{B}_x)^2\le\langle A_x\rangle^2+S^{+}_{xy} \hbox{~~and~~}
(D^{A}_y)^2\le\langle B_y\rangle^2+S^{+}_{xy}.
\label{eq: Condition 2}
\end{equation}
These hold for every $x$ and $y$. Note that the simultaneous saturation of
these eight inequalities requires that $\sin^2 2\chi\!=\!S^{+}_{xy}$ for every
$x$ and $y$, while cases such as
$\sin^2 2\chi\!=\!S^{+}_{00}\!=\!S^{+}_{01}\!=\!S^{-}_{10}\!=\!S^{+}_{11}$
frequently occur in general two-qubit realizations.
We have compared Eq.\ (\ref{eq: Condition 2}) with the corresponding
device-independent bound obtained numerically (by the random tests as used in
Fig.\ \ref{fig: Sufficient} below). The results indicate
that, {\it for two-qubit realizations saturating both
Eqs.\ (\ref{eq: Condition 1}) and
(\ref{eq: Condition 2})}, the two bounds agree with each other within numerical
accuracy, as expected. Moreover, it is found that any correlation, whose
(non two-qubit) realization saturates both Eqs.\ (\ref{eq: Condition 1}) and
(\ref{eq: Condition 2}), and fulfills one more condition
\begin{equation}
\prod_{xy}[(1-S^{+}_{xy})\langle A_x B_y\rangle-\langle A_x\rangle\langle B_y\rangle]\ge 0,
\label{eq: Condition 3}
\end{equation}
always has a two-qubit realization
(see Appendix \ref{sec: Two-qubit realization}). Note that
Eq.\ (\ref{eq: Condition 3}) is merely redundant, when two-qubit realizations
only are considered.

Therefore, the necessary and sufficient condition we propose for the
extremality is the simultaneous saturation of the two inequalities given by
Eq.\ (\ref{eq: Condition 1}) and the eight inequalities given by
Eq.\ (\ref{eq: Condition 2}), and fulfillment of Eq.\ (\ref{eq: Condition 3}).
To check the validity, it suffices to investigate two-qubit realizations,
because of the existence of a two-qubit
realization due to \cite{Tsirelson80a,Masanes06a} and
Eq.\ (\ref{eq: Condition 3}), and the certifiability of $D^{B}_x$
and $D^{A}_y$ already confirmed numerically.
We have performed numerical calculations
to check the necessity of the proposed extremal condition as follows:
For a randomly constructed Bell expression Eq.\ (\ref{eq: Bell expression}),
where without loss of generality all coefficients are randomly selected from 
$[-1,1]$, a two-qubit realization that maximizes the expression is obtained
via the seesaw iteration algorithm \cite{Werner01b,Vertesi14a} using the
semidefinite programming package \cite{SDPA}. A correlation picked up in
this way using a random Bell expression (a random hyperplane in the probability
space) could be a point on a non-extremal boundary,
if the hyperplane were precisely parallel to the boundary. However,
such a coincidence is quite rare in the random tests; hence a point picked up
is in practice always an extremal point. For the same reason, the random tests
cannot pick up a given extremal point unless it has infinite supporting
hyperplanes. However, an implicit continuity assumption for the continuous
distribution of the other extremal points justifies this methodology
(see, e.g., Fig.\ 1 in \cite{Goh18a}).

\begin{figure}[t]
\centerline{\scalebox{0.5}[0.5]{\includegraphics{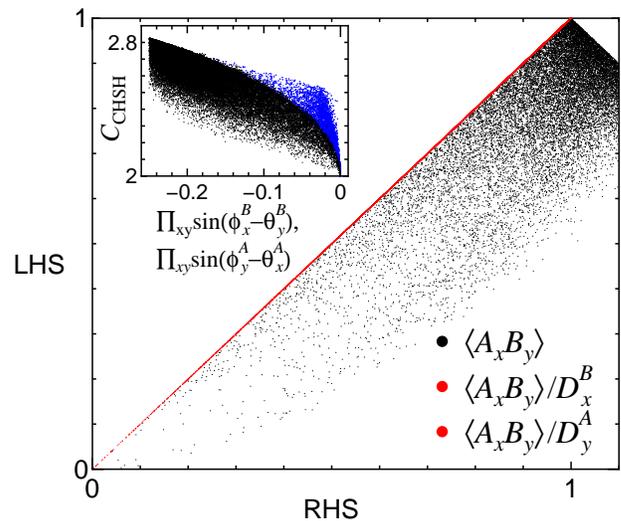}}}
\caption{
The relation between the RHS and LHS of Eq.\ (\ref{eq: Condition 1})
for 20,000 randomly chosen nonlocal realizations ($C_{\rm CHSH}\!>\!2$),
when $\langle A_xB_y\rangle$ are not scaled (black dots) and are
scaled by $D^{B}_x$ or $D^{A}_y$ (red dots). The fraction of nonlocal
realizations among all realizations is roughly 0.6\%. In the inset,
both $(\prod_{xy} \sin(\phi^{B}_x\!-\!\theta^{B}_y),C_{\rm CHSH})$ and
$(\prod_{xy} \sin(\phi^{A}_y\!-\!\theta^{A}_x),C_{\rm CHSH})$ are plotted
for the same realizations (black dots).
The blue dots show the results, where
$V_{11}$ is randomly selected from $[-0.05,0.05]$, and $V^{A}_x$, $V^{B}_y$
are from $[-0.0125,0.0125]$ (the others are from $[-1,1]$), to effectively
pick up rare cases.
}
\label{fig: TLM-type}
\end{figure}

For such realizations obtained randomly,
Fig.\ \ref{fig: TLM-type} shows the relation between the left-hand side (LHS)
and the right-hand side (RHS) of Eq.\ (\ref{eq: Condition 1}). 
When the correlation functions are not scaled
by either $D^{A}_x$ or $D^{B}_y$, the LHS is less than or equal to the RHS, as
shown by the black dots, which is an expected behavior of the (non-scaled) TLM
inequality. However, as shown by the red dots, the equality holds when the
correlation functions are scaled by $D^{A}_x$ or $D^{B}_y$, hence suggesting
that the saturation of Eq.\ (\ref{eq: Condition 1}) is indeed necessary for
the extremality. Note that the saturation is equivalent to the fulfillment
of (see Appendix \ref{sec: Two-qubit realization})
\begin{equation}
\prod_{xy}\sin(\phi^{B}_x\!-\!\theta^{B}_y)\le 0 \hbox{~~and~~}
\prod_{xy}\sin(\phi^{A}_y\!-\!\theta^{A}_x)\le 0,
\label{eq: Condition 1x}
\end{equation}
which is numerically more feasible for verifying the saturation of
Eq.\ (\ref{eq: Condition 1}) than checking Eq.\ (\ref{eq: Condition 1}) itself.
Here, $\phi^{B}_x$ is the angle of the Bloch vector of $\rho^{B}_x$,
$\theta^{B}_y$ is the angle of the measurement basis of $B_y$, and similarly
for Alice.
The inset in Fig.\ \ref{fig: TLM-type} shows the distributions of
$C_{\rm CHSH}$, $\prod_{xy} \sin(\phi^{B}_x\!-\!\theta^{B}_y)$, and
$\prod_{xy} \sin(\phi^{A}_y\!-\!\theta^{A}_x)$ for realizations chosen
randomly. The results indicate that Eq.\ (\ref{eq: Condition 1x}) is
satisfied for all extremal correlations, which strengthens the
results of the main body of Fig.\ \ref{fig: TLM-type}.

We have also performed similar numerical calculations and confirmed that
$(D^{B}_x)^2\!-\!\langle A_x\rangle^2$ and 
$(D^{A}_y)^2\!-\!\langle B_y\rangle^2$
are closer to $S^{+}_{xy}$ than $S^{-}_{xy}$ for every $x$ and $y$ and for all
extremal correlations chosen randomly.
Namely, the numerical results suggest that the saturation of
Eq.\ (\ref{eq: Condition 2}) is also necessary for the extremality.

Let us then investigate the sufficiency. We present the numerical evidence that
a correlation generated by a two-qubit realization, which saturates both
Eqs.\ (\ref{eq: Condition 1}) and (\ref{eq: Condition 2}), is always located at
a quantum boundary. In the calculations, we randomly
construct a realization by
selecting $\theta^{A}_x$, $\theta^{B}_y$, and $\chi$ uniformly.
The realization is discarded if it does not satisfy
Eq.\ (\ref{eq: Condition 1x}). Otherwise, it is kept, and 
\begin{equation*}
\Delta=\max_{xy}\{\langle A_x \rangle^2+S^{+}_{xy}-(D^{B}_{x})^2,
\langle B_y \rangle^2+S^{+}_{xy}-(D^{A}_{y})^2\}
\end{equation*}
is calculated. The realization constructed in this way saturates both
Eqs.\ (\ref{eq: Condition 1}) and (\ref{eq: Condition 2}) only when
$\Delta\!=\!0$. Letting $\mathbf{p}$ be the correlation generated by
the realization, we then investigate the quantum realizability of 
$\mathbf{q}\!=\!\lambda \mathbf{p}+(1-\lambda)\mathbf{I}$,
where $\mathbf{I}$ is the completely random correlation given by
$\langle A_xB_y\rangle\!=\!\langle A_x\rangle\!=\!\langle B_y\rangle\!=\!0$.
Concretely, we obtain the maximum possible value of $\lambda$, denoted by
$\lambda_{\rm max}$, using the $1\!+\!AB$ level of the NPA hierarchy
method for each of the realizations constructed randomly (including the case of
$\Delta\!\ne\!0$). Since $\lambda_{\rm max}$ obtained via the NPA method is
an upper bound such that $\mathbf{q}$ is quantum realizable but $\mathbf{p}$ is
known to be quantum realizable, $\lambda_{\rm max}\!=\!1$ means that
$\mathbf{p}$ is located at a quantum boundary (see the schematic picture
in Fig.\ \ref{fig: Sufficient}). Figure \ref{fig: Sufficient} shows the results
of the calculations, which indicate that $\lambda_{\rm max}\!=\!1$ always
holds when $\Delta\!=\!0$. We have also confirmed that all data points with
$\lambda_{\rm max}\!=\!1$ for $\Delta\!>\!0$ correspond to the edge of the
probability space [$\min p(ab|xy)\!=\!0$].
Note that the device-independent upper bounds of $D^{B}_x$ and $D^{A}_y$ are
typically monotonically decreasing in $\lambda$, while the lower
bounds in Eq.\ (\ref{eq: CQB}) is monotonically increasing \cite{Ishizaka17a}.
These monotonicities also suggest that ${\mathbf p}$, where the two bounds
meet, must be located at a quantum boundary.

\begin{figure}[t]
\centerline{\scalebox{0.5}[0.5]{\includegraphics{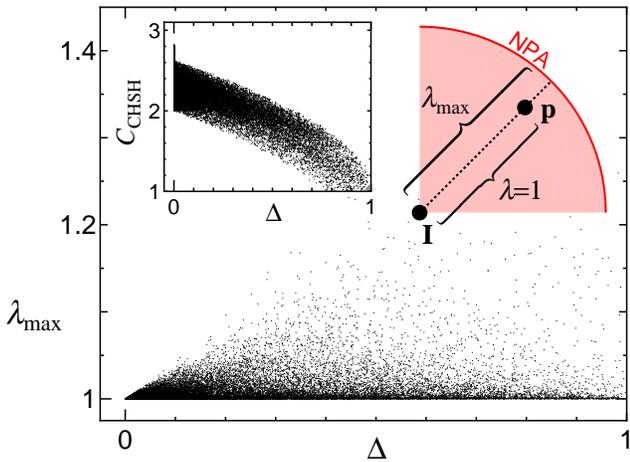}}}
\caption{
For each of 40,000 randomly constructed realizations, $\lambda_{\rm max}$ is
obtained using the NPA hierarchy, and $(\Delta,\lambda_{\rm max})$
is plotted. The inset shows the distribution of $C_{\rm CHSH}$ and $\Delta$,
which indicates that $C_{\rm CHSH}\!\ge\!2$ when $\Delta\!=\!0$.
}
\label{fig: Sufficient}
\end{figure}

Unfortunately, however, the above calculations cannot exclude the possibility
that ${\mathbf p}$ is located at a non-extremal boundary. In the first place,
does there exist any two-qubit realization that can generate such a
non-extremal (and nonlocal) boundary correlation?
This alone is an intriguing but difficult problem as discussed
in \cite{Donohue15a,Goh16a}. In the case of a correlation whose two-qubit
realization saturates both Eqs.\ (\ref{eq: Condition 1}) and
(\ref{eq: Condition 2}), however, the certifiability of
${\mathbf D}\!\equiv\!(D^{B}_0,D^{B}_1,D^{A}_0,D^{A}_1)$
(confirmed numerically) strongly constrains the possibility of being
a non-extremal boundary correlation. For a correlation written by two
extremal correlations as
$\lambda{\mathbf p_0}\!+\!(1\!-\!\lambda){\mathbf p_1}$,
a realization with 
$\sqrt{\lambda{\mathbf D}^2({\mathbf p_0})\!+\!(1\!-\!\lambda){\mathbf D}^2({\mathbf p_1})}$
necessarily exists \cite{Ishizaka17a}, but it must coincide with
${\mathbf D}(\lambda{\mathbf p_0}\!+\!(1-\lambda){\mathbf p_1})$ so that
${\mathbf D}$ is unique. This coincidence is quite unlikely due to the
nonlinear characteristics of the bounds Eq.\ (\ref{eq: Condition 2}),
unless the bounds are constant over the entirety of the boundary.
For example, Fig.\ \ref{fig: Nonextreme} plots the bounds
along the non-extremal boundary illustrated in the figure, where 
the nonlinearity indeed prevents the coincidence.
Figure \ref{fig: Nonextreme} also indicates that 
the disagreement between $S^{+}_{11}$ and
$S^{+}_{00}\!=\!S^{+}_{01}\!=\!S^{+}_{10}$ prevents
the simultaneous saturation of Eq.\ (\ref{eq: Condition 2}).
Note that any two-qubit realization does not exist on the middle of this
boundary as proved in \cite{Donohue15a}. In the case of local correlations,
however, there exist non-extremal boundaries such that the bounds
Eq.\ (\ref{eq: Condition 2}) are constant. For example,
the local non-extremal boundary correlations 
\begin{equation}
\langle A_x \rangle\!=\!\langle B_y \rangle\!=\!0,\hbox{~}
\langle A_0B_y \rangle\!=\!1,\hbox{~}
\langle A_1B_y \rangle\!=\!\lambda,
\label{eq: local nonextremal}
\end{equation}
saturate both Eqs.\ (\ref{eq: Condition 1}) and
(\ref{eq: Condition 2}), where ${\mathbf D}\!=\!(1,1,1,1)$ \cite{Goh18a}.
However, such a non-extremal boundary
(i.e., where the bounds Eq.\ (\ref{eq: Condition 2}) become constant) is
also unlikely, except for ${\mathbf D}\!=\!(1,1,1,1)$, due to the nonlinearity.
These observations combined with the numerical results (and the initial insight
regarding self-testing) motivate us to make the following conjecture:

\begin{figure}[t]
\centerline{\scalebox{0.45}[0.45]{\includegraphics{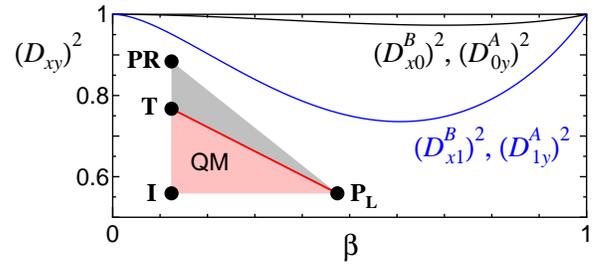}}}
\caption{
In the correlation space 
$\alpha\mathbf{PR}\!+\!\beta\mathbf{P_L}\!+\!(1\!-\!\alpha\!-\!\beta)\mathbf{I}$,
the quantum boundary is the red straight
line as proved in \protect\cite{Goh18a}.
Here, $\mathbf{PR}$ is the postquantum
correlation produced by the Popescu-Rohrlich box \protect\cite{Popescu94a},
$\mathbf{P_L}$ is a local deterministic correlation with
$\langle A_x\rangle\!=\!\langle B_y\rangle\!=\!1$, and
$\mathbf{T}$ is a correlation attaining the Tsirelson bound.
Along this boundary, 
$(D^{B}_{xy})^2\!\equiv\!\langle A_x\rangle^2\!+\!S^{+}_{xy}$ and
$(D^{A}_{xy})^2\!\equiv\!\langle B_y\rangle^2\!+\!S^{+}_{xy}$
are plotted.
Since $D^{B}_x\!=\!D^{A}_y\!=\!1$ at both edges of the boundary,
the correct device-independent upper bounds are equal to 1 over the entirety of
the boundary.
}
\label{fig: Nonextreme}
\end{figure}

{\bf Conjecture 1.}
{\it In nonlocal quantum correlations, a correlation is extremal if and only if
it fulfills Eq.\ (\ref{eq: Condition 3}) and the realization saturates
Eqs.\ (\ref{eq: Condition 1}) and (\ref{eq: Condition 2}).
}

Note that, in the case of unbiased marginals, the saturation of
Eq.\ (\ref{eq: Condition 2}) implies
$D^{B}_x\!=\!D^{A}_y\!=\!1$, and
Eq.\ (\ref{eq: Condition 1}) is reduced to the TLM inequality, as it should be.
Our conjectured criterion also correctly identifies the analytical examples of
extremal correlations in \cite{Acin12a,Ramanathan15a}, and even the non-exposed
extremal correlation of the Hardy point \cite{Goh18a,Rabelo12a}.
Moreover, in the case of local correlations, the inset of
Fig.\ \ref{fig: Sufficient} suggests that the criterion ensures the
boundaryness ($C_{\rm CHSH}\!=\!2$), but not necessarily extremality due to
the correlation of, e.g., Eq.\ (\ref{eq: local nonextremal}). Note further that
Fig.\ \ref{fig: Sufficient} also suggests the following.

{\bf Conjecture 2.}
{\it
The $1\!+\!AB$ level of the NPA
hierarchy (i.e., almost quantumness \cite{Navascues15a}) is sufficiently
strong to tightly bound every extremal correlation.
}

It is immediately noticed that we can eliminate $D^{B}_x$ and $D^{A}_y$ 
by combining Eqs.\ (\ref{eq: Condition 1}) and (\ref{eq: Condition 2}).
The resultant set of inequalities must suffice for identifying the
extremality by virtue of the certifiability of $D^{B}_x$ and $D^{A}_y$.
However, separating the extremal condition into
Eqs.\ (\ref{eq: Condition 1}) and (\ref{eq: Condition 2}) will be advantageous
in searching for fundamental principles that limit nonlocal
correlations, as the principles leading to Eqs.\ (\ref{eq: Condition 1}) and
(\ref{eq: Condition 2}) will be independent of each other.
For example, the information causality (IC) principle \cite{Pawlowski09a}
successfully explains the Tsirelson bound and even some curved quantum
boundaries \cite{Allcock09a}, and it was expected that the IC principle
could explain every quantum boundary.
As noted in \cite{Ishizaka17a}
(see also Appendix \ref{sec: Insufficiency of IC principle}), however, the IC
principle cannot explain extremal boundaries generated from a partially
entangled state, as it cannot explain the saturation of
Eq.\ (\ref{eq: Condition 1}) unless $D^{B}_x\!=\!D^{A}_y\!=\!1$, 
where Eq.\ (\ref{eq: Condition 2}) plays no role, i.e., the IC principle is
unrelated to Eq.\ (\ref{eq: Condition 2}). On the other hand,
the cryptographic principle possibly explains the saturation of
Eq.\ (\ref{eq: Condition 1}) when
$D^{B}_x,D^{A}_y\!<\!1$ \cite{Ishizaka17a},
but it cannot explain Eq.\ (\ref{eq: Condition 2}).

What is the fundamental principle that leads to Eq.\ (\ref{eq: Condition 2})? 
This is an important open problem, but the physical meaning of
Eq.\ (\ref{eq: Dx and Dy}), on which Eq.\ (\ref{eq: Condition 2}) is based, is
relatively obvious: the entanglement bound for the guessing probability in an
uncertainty game \cite{Berta10a,Coles17a}. When Alice and Bob share a maximally
entangled state ($\sin^2 2\chi\!=\!1$), Bob can perfectly guess Alice's outcome
for both $x\!=\!0,1$ as $D^{B}_x\!=\!1$, and the uncertainty between $A_0$ and
$A_1$ vanishes \cite{Einstein35a}. The guessing probability decreases as the
entanglement decreases, and for an unentangled state
($\sin^2 2\chi\!=\!0$), the guessing probability is solely determined by the
uncertainty $\Delta A_x\!=\!\sqrt{1\!-\!\langle A_x \rangle^2}$ as
$D^{B}_x\!=\!|\langle A_x\rangle|$
(see \cite{Oppenheim10a,Ramanathan15a,Zhen16a}
for a slightly different link between nonlocality and
uncertainty). Since correlations with biased marginals are
generated from a partially entangled
state, it is natural that the amount of entanglement is involved in the
extremality condition. Hence, a fundamental principle that leads to
Eq.\ (\ref{eq: Condition 2}) must be the one that more or less explains the
entanglement bound in an information-theoretical way.

The plausible analytical condition that limits the strength of extremal
quantum correlations in the simplest Bell scenario was determined.
We hope that this analytical condition will result in a new fundamental
principle behind quantum mechanics to be found.

The author would like to thank the anonymous referees for pointing out 
Eq. (\ref{eq: local nonextremal}) and for suggestions on improving the
presentation of this paper.
This work was supported by JSPS KAKENHI Grant No. 17K05579. 

%

\appendix

%
\section{Two-qubit realization}
\label{sec: Two-qubit realization}

Here, the details of two-qubit realizations, where projective measurements
of rank 1 are performed on a two-qubit entangled state
$|\psi\rangle$, are described.
By applying appropriate local unitary transformations, Alice's and Bob's
observables are written as
\begin{equation}
A_x=\cos \theta^{A}_x \sigma_1 + \sin \theta^{A}_x \sigma_3,\hbox{~}
B_y=\cos \theta^{B}_y \sigma_1 + \sin \theta^{B}_y \sigma_3,
\label{eq: Parameterization 1}
\end{equation}
where $(\sigma_1,\sigma_2,\sigma_3)$ are the Pauli matrices. Since 
any Bell expression is then maximized when
$\rho\!=\!|\psi\rangle\langle\psi|$ is real symmetric,
$|\psi\rangle$ can be expressed by further rotating the local bases as
\begin{equation}
|\psi\rangle=\cos\chi|00\rangle+\sin\chi|11\rangle
\hbox{~~~($0\!<\!\chi\!\le\!\pi/4$)}.
\end{equation}
Under this parameterization, we have
\begin{eqnarray}
\langle A_xB_y\rangle&=& \sin\theta^{A}_x\sin\theta^{B}_y+\cos\theta^{A}_x\cos\theta^{B}_y \sin2\chi,
\label{eq: Cxy}  \\
\langle A_{x}\rangle&=& \sin\theta^{A}_x \cos2\chi,
\label{eq: Ax}  \\
\langle B_{y}\rangle&=& \sin\theta^{B}_y \cos2\chi.
\label{eq: By}
\end{eqnarray}
Moreover, define the angles $\phi^{B}_x$ and $\phi^{A}_y$ as
\begin{eqnarray}
\hbox{tr}_A A_x |\psi\rangle\langle\psi|
&\!\!\!=\!\!\!&
\frac{\langle A_x\rangle}{2} I +
\frac{D^{B}_x}{2}(\cos \phi^{B}_x \sigma_1 + \sin \phi^{B}_x \sigma_3), \cr
\hbox{tr}_B B_y |\psi\rangle\langle\psi|
&\!\!\!=\!\!\!&\frac{\langle B_y\rangle}{2} I + 
\frac{D^{A}_y}{2}(\cos \phi^{A}_y \sigma_1 + \sin \phi^{A}_y \sigma_3).
\label{eq: Parameterization 3}
\end{eqnarray}
It is found that
\begin{eqnarray}
D^{B}_x &\!\!\!=\!\!\!&\sqrt{\sin^2 \theta^{A}_x+\cos^2 \theta^{A}_x\sin^2 2\chi}
=\hbox{tr}|\rho^{B}_{1|x}\!-\!\rho^{B}_{-1|x}|, \cr
D^{A}_y &\!\!\!=\!\!\!&\sqrt{\sin^2 \theta^{B}_y+\cos^2 \theta^{B}_y\sin^2 2\chi}
=\hbox{tr}|\rho^{A}_{1|y}\!-\!\rho^{A}_{-1|y}|,
\end{eqnarray}
where
$\rho^{B}_{a|x}\!=\!\hbox{tr}_A\frac{I+aA_x}{2}|\psi\rangle\langle\psi|$
and
$\rho^{A}_{b|y}\!=\!\hbox{tr}_B\frac{I+bB_y}{2}|\psi\rangle\langle\psi|$.

Let us then determine the entanglement specified by $\sin^22\chi$ for a given
$\{\langle A_xB_y\rangle,\langle A_x\rangle,\langle B_y\rangle\}$.
Eliminating $\theta^{A}_x$ and $\theta^{B}_y$ from
Eqs.\ (\ref{eq: Cxy})\textendash(\ref{eq: By}),
we have 
\begin{equation*}
\langle A_x B_y\rangle=\frac{\langle A_x\rangle\langle B_y\rangle}{\cos^22\chi}
\pm\sin2\chi
\sqrt{1-\frac{\langle A_x\rangle^2}{\cos^22\chi}}
\sqrt{1-\frac{\langle B_y\rangle^2}{\cos^22\chi}},
\end{equation*}
and thus we have the quadratic equation for $\cos^22\chi$, i.e.,
\begin{equation}
\cos^42\chi + (J_{xy}-2)\cos^22\chi+K^{2}_{xy}-J_{xy}+1=0.
\label{eq: quadratic}
\end{equation}
Since this must hold for every $x$ and $y$, there are four quadratic
equations in total. Two solutions of each quadratic equation are given by
$\sin^22\chi\!=\!S^{\pm}_{xy}$.

Let us see that, when a (non two-qubit) realization of a correlation
simultaneously saturates the scaled TLM
inequalities,
$(D^{B}_x)^2\!\le\!\langle A_x\rangle^2\!+\!S^{+}_{xy}$,
$(D^{A}_y)^2\!\le\!\langle B_y\rangle^2\!+\!S^{+}_{xy}$,
and further fulfills
$\prod_{xy}[(1-S^{+}_{xy})\langle A_xB_y\rangle\!-\!\langle A_x\rangle\langle B_y\rangle]\!\ge\!0$,
the correlation also has a two-qubit realization.
When $S^{+}_{xy}\!=\!1$, since $D^{B}_x,D^{A}_y\!\le\!1$, it is found that
$D^{B}_x\!=\!D^{A}_y\!=\!1$ in the original realization, and the existence of a
two-qubit realization is obvious from the TLM criterion.
When $0\!<\!S^{+}_{xy}\!<\!1$, a two-qubit realization can be constructed as
follows: first determine $\chi$ from $\sin^22\chi\!=\!S^{+}_{xy}$, and next
determine $\sin\theta^{A}_x$ and $\sin\theta^{B}_y$ from
Eq.\ (\ref{eq: Ax}) and Eq.\ (\ref{eq: By}), respectively. This two-qubit
realization can reproduce $\langle A_xB_y\rangle$ of the original realization
by adjusting the signs of $\cos\theta^{A}_x$ and $\cos\theta^{B}_y$,
as $S^{+}_{xy}$ is a solution of Eq.\ (\ref{eq: quadratic}), and
$\prod_{xy}[\langle A_x B_y\rangle\!-\!\frac{\langle A_x\rangle\langle B_y\rangle}{\cos^22\chi}]\ge 0$.
Moreover, since $D^{B}_x$ and $D^{A}_y$ of the two-qubit realization are the
same as those of the original realization, the two-qubit realization saturates
the scaled TLM inequality, if the original realization does.

Note that
\begin{eqnarray}
\sin\theta^{A}_x &\!\!=\!\!& D^{B}_x \sin\phi^{B}_x, \hbox{~~}
\cos\theta^{A}_x = D^{B}_x \cos\phi^{B}_x/\sin2\chi , \nonumber\\ 
\sin\theta^{B}_y &\!\!=\!\!& D^{A}_y \sin\phi^{A}_y, \hbox{~~}
\cos\theta^{B}_y = D^{A}_y \cos\phi^{A}_y/\sin2\chi ,
\end{eqnarray}
hence,
\begin{equation}
\frac{\langle A_x B_y\rangle}{D^{B}_x}=\cos(\phi^{B}_x-\theta^{B}_y),\hbox{~}
\frac{\langle A_x B_y\rangle}{D^{A}_y}=\cos(\phi^{A}_y-\theta^{A}_x).
\end{equation}
When $\tilde C_{xy}\!=\!\cos(\phi_x\!-\!\theta_y)\!\equiv\!\cos\delta_{xy}$,
by noticing that
\begin{eqnarray*}
\tilde C_{00}\tilde C_{01}-\tilde C_{10}\tilde C_{11}&=&
\cos\delta_{00}\cos\delta_{01}-\cos\delta_{10}\cos\delta_{11} \cr
&=&
\cos(\delta_{00}-\delta_{01})-\cos(\delta_{10}-\delta_{11}) \cr
&&-\sin\delta_{00}\sin\delta_{01}+\sin\delta_{10}\sin\delta_{11} \cr
&=&-\sin\delta_{00}\sin\delta_{01}+\sin\delta_{10}\sin\delta_{11},
\end{eqnarray*}
it is not difficult to see that the necessary and sufficient condition
for the saturation of the scaled TLM inequality is given by
$\sin\delta_{00}\sin\delta_{01}\sin\delta_{10}\sin\delta_{11}\!\le\!0$.

%
\section{Insufficiency of IC principle}
\label{sec: Insufficiency of IC principle}

We briefly noted in \cite{Ishizaka17a} that the information
causality (IC) principle is insufficient for the full identification of the
quantum boundaries for bipartite settings, no matter what protocol is
considered. However, the paper was criticized because the explanation was
considered unclear or the point was completely misunderstood.
Here, we explain the point in more detail.

Let us recall the derivation of the IC principle. In the general setting of
communication, where Alice is given a bit string
$\vec x\!=\!(x_1,x_2,\cdots)$ and sends $\vec m$ to Bob as a message, the
information about $\vec x$ obtainable
by Bob is characterized by the mutual information $I(\vec x\!:\!\vec m\rho_B)$,
where $\rho_B$ is the state of Bob's half of the no-signaling resources.
Using the
no-signaling condition and the information-theoretical relations respected by
quantum mechanics, it was shown in \cite{Pawlowski09a} that 
\begin{equation}
I(\vec x:\vec m \rho_B) 
\le H(\vec m)-H(\vec m|\vec x \rho_B) \le H(\vec m).
\label{eq: IC principle}
\end{equation}
Since the entropy $H(\vec m)$ cannot exceed the number of bits in $\vec m$,
the IC principle is derived. Here, we consider the case where the number of
message bits is finite such that $H(\vec m)$ is finite.

\begin{figure}[t]
\centerline{\scalebox{0.45}[0.45]{\includegraphics{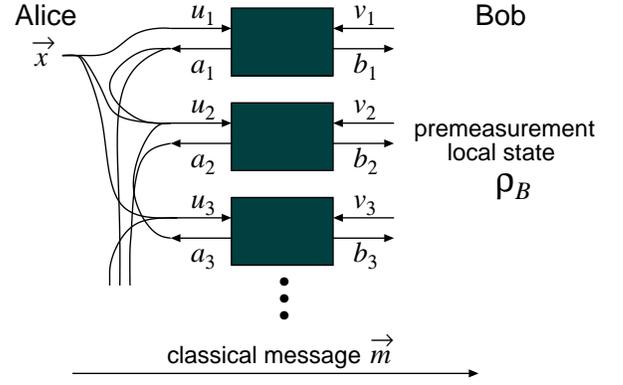}}}
\caption{
A protocol may connect the quantum boxes in a complicated way; regardless,
let us denote the outputs of the boxes as $\vec a\!=\!(a_1,a_2,\cdots,a_n)$.
To achieve the maximum limit set by the IC principle, the protocol must satisfy
$H(\vec m|\vec x\rho_B)\!=\!0$, i.e., Bob must be able to completely determine
$\vec m$ from $\vec x$ and his local state $\rho_B$. The message $\vec m$ is
constructed from $\vec a$ and $\vec x$, but $\vec a$ is ambiguous for Bob. Can
Alice construct an unambiguous message $\vec m$ by using the ambiguous output
$\vec a$ from the quantum boxes?
}
\label{fig: Protocol}
\end{figure}

The point is that the term $H(\vec m|\vec x\rho_B)$
in Eq.\ (\ref{eq: IC principle}) is inevitably nonzero in some cases; hence,
the saturation of Eq.\ (\ref{eq: IC principle}) is impossible. Namely,
the IC principle has omitted the nonnegligible term in its derivation.

Suppose that Alice and Bob share $n$ identical ``quantum boxes'', each of
which accepts inputs $(u,v)$ and produces outputs $(a,b)$ according to
the conditional probabilities $p(ab|uv)$, where the simplest Bell scenario is
considered (see Fig.\ \ref{fig: Protocol}).
A protocol may connect the inputs and outputs of the $n$ boxes in a complicated
way, but let us denote Alice's outcomes as
$\vec a\!=\!(a_1,a_2,\cdots,a_n)$, where $a_i$ is the outcome of the $i$-th
quantum box. Now, consider a correlation located at an extremal boundary and
showing $D^{B}_{u}\!<\!1$. This means that Bob's local states (of a single
box) for different values of $a$ become nonorthogonal; thus, he cannot
completely determine $a$. Since $D^{B}_{u}$ is generally upper bounded in a
device-independent way, his ambiguity about $a$ is
inevitable, irrespective of the details of the realization of the quantum box.
In this way, each $a_i$ has ambiguity for Bob.
This ambiguity is so strong that he cannot determine $\vec a$ even if he knows
$\vec x$ (and even if he knows all of Alice's inputs $\vec u$ to
the boxes), i.e., $H(\vec a|\vec x\rho_B)\!>\!0$. Since $\vec m$ is constructed
from $\vec x$ and $\vec a$, it is clear from Eq.\ (\ref{eq: IC principle}) that
any protocol whose $\vec m$ contains the information of $\vec a$ and
$H(\vec m|\vec x\rho_B)\!>\!0$ cannot achieve
$I(\vec x\!:\!\vec m\rho_B)\!=\!H(\vec m)$.

Note that no redundant coding technique can reduce Bob's ambiguity about
$\vec a$, as the ambiguity originates from $\rho_B$, which is not
under Alice's control. For example, $\vec a\vec a\vec a$ has exactly the same
ambiguity as $\vec a$ for Bob. If Alice postselects the boxes
with the same output $a$ to multiply Bob's local state such as
$\rho^{B}_{a|u}\!\otimes\!\rho^{B}_{a|u}
\!\otimes\!\rho^{B}_{a|u}\!\otimes\!\cdots$,
she can reduce Bob's ambiguity, but such postselection is not allowed.
Although Alice can control the value of $a$ via the
input $u$ to some degree, $a$ is nevertheless determined in a
probabilistic way by $p(a|u)$, and it is impossible to completely
eliminate Bob's ambiguity about $a$.

More concretely, let us consider the quantum box that is realized by a pure
partially entangled state and showing $D^{B}_0,D^{B}_1\!<\!1$. Namely, the
outcome $a$ is ambiguous for both $u\!=\!0$ and $1$, and all $a_i$'s are always
ambiguous for Bob. The only way for $H(\vec m|\vec x\rho_B)\!=\!0$ is that
$\vec m$ does not contain any information of $\vec a$ at all.
This is because, since $\vec m$ is constructed from $\vec a$ and
$\vec x$, and thus $H(\vec m|\vec a\vec x\rho_B)\!=\!0$,
we have $H(\vec m|\vec x\rho_B)\!=\!
H(\vec m|\vec x\rho_B)\!-\!H(\vec m|\vec a\vec x\rho_B)
\!=\!H(\vec a|\vec x\rho_B)\!-\!H(\vec a|\vec m\vec x\rho_B)
\!=\!I(\vec a\!:\!\vec m|\vec x\rho_B)$.
Namely, $H(\vec m|\vec x\rho_B)\!=\!0$ implies 
$I(\vec a\!:\!\vec m|\vec x\rho_B)\!=\!0$;
any information about $\vec a$ must not be obtained via $\vec m$.
In this case, the achievement of
$I(\vec x\!:\!\vec m\rho_B)\!=\!H(\vec m)$ may be possible (the IC
inequality can be saturated even in a purely classical case \cite{Al-Safi11a}),
but the protocol does not utilize the quantum correlation at all. This
protocol, of course, cannot explain the outperformance of the pure entangled
state at all (recall that every pure entangled state violates some Bell
inequality \cite{Gisin91a}), and hence cannot explain the corresponding
extremal boundary.

This is the case of the quantum box showing $D^{B}_0\!<\!1$ but
$D^{B}_1\!=\!1$. Namely, the outcome $a$ is ambiguous only when $u\!=\!0$. To
achieve $H(\vec m|\vec x\rho_B)\!=\!0$, the protocol can only utilize the
unambiguous outcomes when $u\!=\!1$. In this case, however,
without changing the performance of the protocol, we can replace all $a_i$'s
corresponding to $u_i\!=\!0$ with a fixed value, e.g., $1$ (because these are
not used), and the correlation produced by the quantum boxes is replaced with
the classical one.
This implies that the behavior of the protocol can be simulated exactly using
classical correlations; hence, this protocol cannot explain the outperformance
of the pure entangled state again.

In the other remaining case where $D^{B}_0\!=\!D^{B}_1\!=\!1$, if the quantum
box is realized by a pure partially entangled state, the produced correlation
is classical, because Alice's measurement bases for $u\!=\!0$ and $1$ both
agree with the Schmidt basis of the pure state.

From the above, it is concluded that the nonlocal extremal boundaries that can
be tightly explained by the IC principle are those realized by a maximally
entangled state, where $D^{B}_0\!=\!D^{B}_1\!=\!D^{A}_0\!=\!D^{A}_1\!=\!1$.


%

\end{document}